# A novel forward-looking ultrasound catheter for treating vascular occlusions

Jingjing Liu, Alex R. Wright, Kullervo Hynynen, and David E. Goertz

ABSTRACT - Thrombotic and chronic occlusions of large blood vessels are a major cause of mortality and morbidity, and so there is a need for improved treatments in many clinical circumstances. Endovascular ultrasound approaches have been shown to hold considerable potential to treat large vessel thrombotic occlusions. Here, we report the development of a novel forward-looking therapeutic ultrasound catheter approach. The design concept centers on the use of a radially polarized hollow cylindrical transducer situated at the distal tip. This approach enables a compact configuration where the central lumen can accommodate a guidewire during navigation as well as provide a route to release cavitation seeds and therapeutic agents adjacent to the site of occlusion. PZT-5H transducers with outer/inner diameters of 1.35/0.73 mm were evaluated using simulations and experiments for their capacity to project forward-looking ultrasound. A length of 2.5 mm operating in the 3$^{rd}$ harmonic of the length mode resonance (1.85 MHz) was selected. Catheters (1.55 mm outer diameter) were designed and fabricated with a liner, outer jacket, and braiding, where the transducers were incorporated with air-backing and a front-face matching layer. Forward-looking pressures of 2.5 MPa (peak negative) at 0.5 mm were achieved. Proof of principle vessel phantom experiments were performed demonstrating the ability to eject microbubbles in proximity to an occlusion and stimulate inertial cavitation. This approach holds potential for treating thrombotic and chronic total occlusions.

## I. INTRODUCTION

THE occlusion of large blood vessels with thrombus is a major cause of mortality and morbidity worldwide [1], [2], [3]. It occurs in the context of stroke, myocardial infarction, pulmonary embolism, and in the peripheral vasculature. Acute thrombotic occlusions are frequently treated with thrombolytic agents and increasingly catheter-based thrombectomy approaches [4], [5], [6], [7], which have the objective of degrading or removing clots, respectively. Despite recent advances, there remains a considerable need for improved treatments of acute and subacute thrombotic occlusions in many clinical circumstances.

In the coronary and peripheral vasculature, thrombotic occlusions can persist and over time their composition evolves, becoming increasingly invested with collagen and potentially calcifications. These are referred to as chronic total occlusions (CTOs), which can exhibit complex morphologies that vary with age and anatomic location. In the case of coronary arteries, of particular interest in the present study, a prominent CTO feature is the presence of a stiff collagen-dense proximal fibrous cap (PFC) [8], [9], [10]. CTOs are frequently detected on diagnostic coronary angiograms, with incidence rates in the range of 20% in patients with clinically significant coronary artery disease [11]. Coronary CTOs can present considerable treatment challenges, with approaches including bypass surgery or percutaneous (PCI) methods that typically involve 'crossing' the CTO with a stiff guidewire, then performing balloon angioplasty, followed by stenting [12]. Successful percutaneous CTO revascularization can lead to improvements in cardiac function, quality of life, and may increase long-term survival [9], [13], [14], [15], [16]. Unfortunately, technical difficulties associated with CTO crossings limit revascularization success rates to 55–80% [17], [18], [19]. The high failure rates for attempted CTO PCI procedures are frequently associated with an inability to advance a guide wire across the mechanical barrier presented by the PFC [20], [21], [22]. Newer techniques and innovations to overcome this limitation are a high clinical priority [16]. These have included approaches such as ablation [4], [23], advanced crossing techniques [24], [25], [26], and the use of the enzyme collagenase to mechanically degrade the PFC [27], [28], [29]. The latter approach shows considerable potential for improving coronary CTO revascularization, however an obstacle to its widespread clinical adoption is that it requires a two-day process to allow sufficient time for PFC degradation.

There has been extensive work conducted to investigate sonothrombolysis (STL) — the use of ultrasound to resolve thrombotic occlusions — within the context of large vessel thrombotic occlusions. This has included its use in combination with lytic agents, primarily tissue plasminogen activator (tPa), [30], [31], [32], [33] and in some cases accompanied by the use of encapsulated bubbles [34], [35], [36] or droplets [37], [38], [39]. Histotripsy based STL is also being investigated, both in

This work was supported in part by the Canadian Institutes of Health Research under Grant 165849. (Corresponding author: David E. Goertz). Jingjing Liu, David E. Goertz, and Kullervo Hynynen are with the Department of Medical Biophysics, The University of Toronto, Toronto, ON, M5G 1L7, Canada.

Jingjing Liu, Alex R. Wright, David E. Goertz, and Kullervo Hynynen are with the Physical Sciences, Sunnybrook Research Institute, Toronto, ON, M4N3M5, Canada.

the presence and absence of lytic agents [40], [41]. Relatively little work has been conducted to investigate ultrasound approaches to treat CTOs. Of relevance to the present study are previous reports in *ex vivo* CTO phantoms [42] and *in vivo* rabbit models [43] that demonstrated the ability of microbubble-mediated ultrasound to enhance the effects of collagenase therapy in softening CTOs. This approach has the potential to enable collagenase therapy to be a single session clinical procedure, though it would require implementation in catheter form to facilitate its adoption in an interventional cardiology setting.

While the large majority of sonothrombolysis work to date has involved or been motivated by the use of extracorporeal transducers, intravascular approaches are also being developed. The most advanced in terms of clinical use is the EKOS catheter, which is comprised of multiple transducers situated along its distal segment. The basis of its operation is to insert the catheter into a soft thrombotic occlusion, release lytic agents from side ports, and employ low intensity ultrasound to promote clot degradation [44], [45]. This approach typically takes 6-24 hours while it is indwelling in the clot and is not compatible for insertion into stiffer occlusions such as CTOs. A recent series of papers [46], [47], [48], [49] have reported the development of transducers intended for use in catheter-based forward-looking ultrasound configuration for microbubble (and/or droplet) mediated sonothrombolysis. To achieve operating frequencies in the MHz to sub-MHz range (more suitable for microbubble mediated therapy), the approach taken was to use a 'stacked' transducer operating in lateral mode to overcome electrical impedance challenges associated with small aperture transducers [48], [50]. Prototype assemblies within 11F guide catheters, using a side port for microbubble injection, showed encouraging benchtop thrombolysis results [46], [48], [49]. A final notable intravascular approach is to use an external excitation source to induce vibrations of the tips of specialized wires. This approach has been primarily investigated in the setting of heavily calcified CTOs, such as those present in peripheral arteries [51], [52], [53].

In the present work, we propose and investigate the use of a radially polarized hollow cylindrical transducer (Supplementary Fig. 1) as the active element at the tip of a forward-projecting therapeutic ultrasound catheter. This geometry has several primary resonant modes (length, thickness, and radial), each of which is associated with different acoustic emission patterns. These transducers have been previously employed in biomedical ultrasound in the context of transcranial arrays [54] as well as catheter-based implementations for heating [55]. In the latter case, a thickness mode is exploited to preferentially achieve radially outward propagating waves. In the transcranial case, the length mode is exploited to emit preferentially along the transducer axis and the far field waves from an array of such transducers, situated in a hemispherical pattern, are used to create a distant focal region [54]. A key advantage of the transducers used in this setting is that electrical impedance issues associated with driving small elements at lower frequencies via their thickness mode can be overcome by lateral mode (electrodes on inner and outer surface) stimulation at the length resonant frequency, which improves electromechanical conversion efficiency in the forward-looking direction at sub-MHz frequencies [56].

In the setting of catheter-based microbubble mediated treatments of vascular occlusions, this configuration — with its patent lumen — offers several advantages. First, the lumen can accommodate a guidewire which will enable the clinician to navigate the catheter tip to the occlusion's proximal surface. This is of particular relevance for coronary arteries or smaller peripheral vessels as it permits a more compact configuration. Indeed, it parallels the approach taken with 'microcatheters' that are employed for CTO treatments, where a (stiff) guidewire situated within the lumen is employed to indent the occlusion surface during treatments. Second, bubbles (or other cavitation seeds) and therapeutic agents can be introduced through the lumen so that they exit at the site of the occlusion and within the ultrasound beam. This obviates the need for side ports, which can enable a more compact catheter configuration.

In this study, we first use simulations and experiments to investigate the forward-projected field of mm-scale hollow cylindrical transducers and select a length as well as operating mode. Prototype catheters are then developed with the selected element, where backing and matching layers are employed. The catheter employs clinical grade liners, jackets, and braiding. Lastly, the performance of the catheter is characterized and a proof of principle experiment with microbubbles in a channel phantom is performed.

## II. MATERIALS AND METHODS

### A. Transducer elements

The radially polarized cylindrical transducer elements (Morgan Technical Ceramics) were made from PZT-5H [57]. The transducer elements had outer diameters of 1.35 mm and inner diameters of 0.73 mm. A ground electrode covers the outer surface of the cylinders while the inner surface forms the signal electrode connection.

### B. Computer Simulation

Transducer simulations were conducted using the finite element analysis software package OnScale™. The cylindrical geometry was represented as a 2D axisymmetric shape, as seen in Supplementary Fig. 2. Lengths from 1 to 5 mm in 0.5 mm increments were simulated, with the selected meshing having been based on convergence tests. To capture frequency-dependent characteristics of complex impedance and surface displacement, a 10 MHz single cycle sine wave served as the driving function. Surface displacement was measured in the axial and radial directions using a node on the flat face and a node at the midpoint on the outer surface, respectively. Pressure maps were captured using 20-cycle sine waves at the specific mode frequency and L-C matching circuits were implemented in the simulation to match the transducer to 50 Ω.

### C. Experimental

The elements were cut to lengths from 2 to 4 mm in 0.5 mm increments using a diamond dicing saw (Model DAD 3240, Disco Co., Japan). To allow for rapid testing of multiple elements, the electrode connection methods were designed to be non-permanent and repeatable. The cylinder rested on a straight length of 250 µm diameter spring steel wire passing through the inner lumen up to the open face of the cylinder. The ground connection was formed with a similar steel wire positioned outside of the cylinder and running perpendicular to the signal connection (Supplementary Fig. 3). The ground wire was tensioned to provide a light and consistent downward force on the outside of the cylinder which in turn pressed the cylinder onto the inner signal wire, thereby ensuring a reliable electrical connection for both the signal and ground as well as physically securing the cylinder.

Transducer elements were tested in a 3D scanning hydrophone tank filled with degassed deionized water (Fig. 1). Elements were mounted in the tank and their inner lumens were flushed with water to remove any potential trapped gases. Scanning was done using a 10 µm active element fiber optic hydrophone (Precision Acoustics, UK) attached to a computer-controlled 3D stage (Velmex, Inc., USA). The fiber optic hydrophone is not impacted by electrical coupling with the transmit signal that impacts PVDF hydrophones. A PCI digitizer (Model DP310, Acqiris, Switzerland) captured the hydrophone signals. A camera attached to the stage facilitated alignment and positioning of the hydrophone.

The connection to the transducer could be switched between a network analyzer (Model AA-30.ZERO, Rig Expert Ukraine Ltd., Ukraine) and the driver path, which consisted of an in-house adjustable L-C matching circuit, an RF power amplifier (Model A150, Electronics & Innovation, Ltd., USA), and a function generator (Model AFG3022B, Tektronix, USA). The matching circuit used a series of computer-controlled relays to switch inductors and capacitors on or off. This allowed for fine discrete control to dynamically match each element to 50 Ω (± 3 Ω) and 0° (± 5°) across a range of frequencies. All scanning and matching processes were controlled using a custom MATLAB interface.

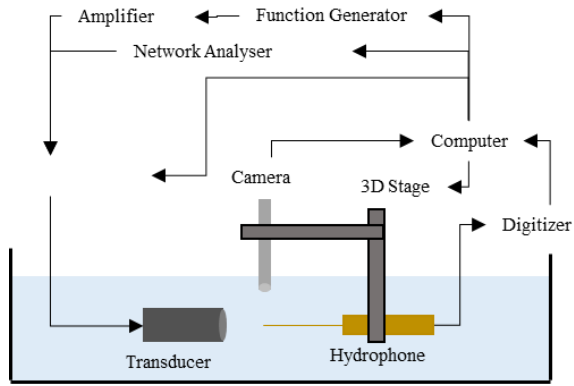

**Fig. 1**. Schematic of hydrophone scanning tank apparatus and component connections.

To select resonant frequencies, the system was configured to perform 2D spatial scans. This created pressure maps showing the spatial distribution outwards from the transducer face for a given frequency and excitation voltage. The system was configured to perform voltage sweeps from 10 to 300 V (peak-to-peak) in steps of 10 V.

### D. Catheter Construction

A schematic overview of the catheter structure is shown in Fig. 2(a). A 40 AWG coaxial cable was attached to the signal and ground electrodes using silver epoxy and strips of copper foil (15 um thickness). The transducer element was placed at the catheter tip between a liner and outer jacket. Thin-walled medical grade PTFE tubing (Streamliner™, Zeus Inc., USA) that matched the PZT cylinder internal diameter was used as the catheter liner (0.05 mm thick). Larger PTFE tubing covered the PZT outer surface and sealed the proximal end of the PZT cylinder to create an air backing layer. A nylon jacket (Pebax, Zeus Inc., USA) was used as the outer jacket on top of the outer PTFE tube to provide structural integrity (~0.25 mm thick). The proximal portion of the catheter had a layer of braided wire between the liner and outer jacket to provide mechanical support and flexibility. For four of the assembled catheters, an epoxy acoustic matching layer (1161M UV Epoxy, Dymax, USA) was deposited on the front face of the transducer and trimmed to the desired thickness with a dicing saw (Model DAD 3240, Disco Co., Japan). The epoxy matching layer was designed to have a thickness of $\lambda_{Epoxy}/4$ at the length mode 3$^{rd}$ harmonic, $f_{L3}$. One catheter was assembled without a matching layer for comparison. The proximal portion of each catheter lumen was connected to a customized injection port to enable the introduction of fluids. A constructed catheter-based transducer is shown in Fig. 2(b).

The catheter-based transducers were characterized according to electrical impedance and pressure profiles were measured according to the previously described methods.

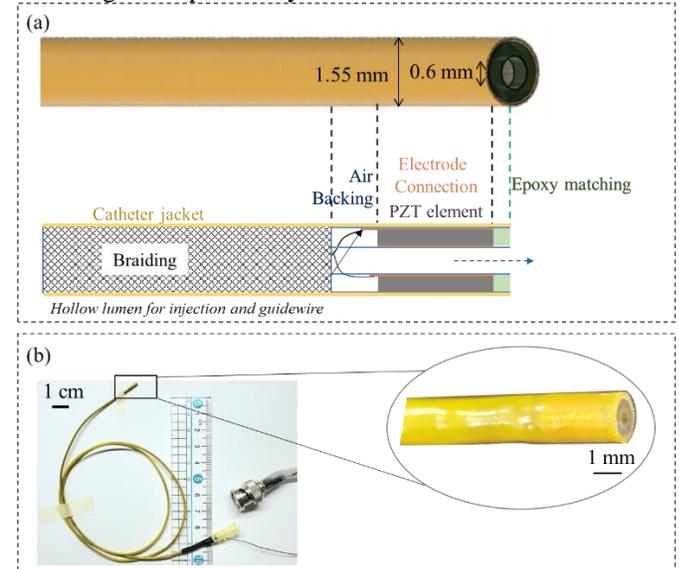

**Fig. 2**. (a) Schematic of the catheter-based transducer structures and (b) a constructed catheter.

### E. Vessel phantom preparation and microbubble infusion

Fig. 3(a) shows the setup for vessel phantom experiments. A vessel channel was cast in agarose (Sigma Aldric, USA) with a diameter of 3 mm and an occlusion. The catheter-based transducer was inserted into the channel phantom such that the tip remained 3 mm away from the occluded surface and the position was fixed with a Tuohy Borst adapter (Cook Medical, USA). A receiver transducer (Panametrics C305, 2.25 MHz, D = 0.75'', F = 1'') was focused onto the channel surface to acquire cavitation signals. A 2 MHz high-pass filter was introduced to reduce the transmission signals at 1.85 MHz. The received signal was then amplified with a 35 dB pre-amplifier. The sonication process was monitored via ultrasound imaging using a 20-MHz probe (MS250D, VisualSonics, USA) on a Vevo 2100 (VisualSonics, USA) at a frame rate of 100 Hz and a transmission power setting of 1%.

The contrast agent Definity™ (Lantheus Medical Imaging, USA) was activated and diluted in PBS at a ratio of 1:1500. The Definity microbubbles were infused from a port at the proximal end of the catheter using a syringe pump (Harvard Apparatus, USA). The user-defined pumping sequences were controlled by MATLAB and coordinated with the sonication pulses. The pulsing parameters used are listed in Fig. 3(b).

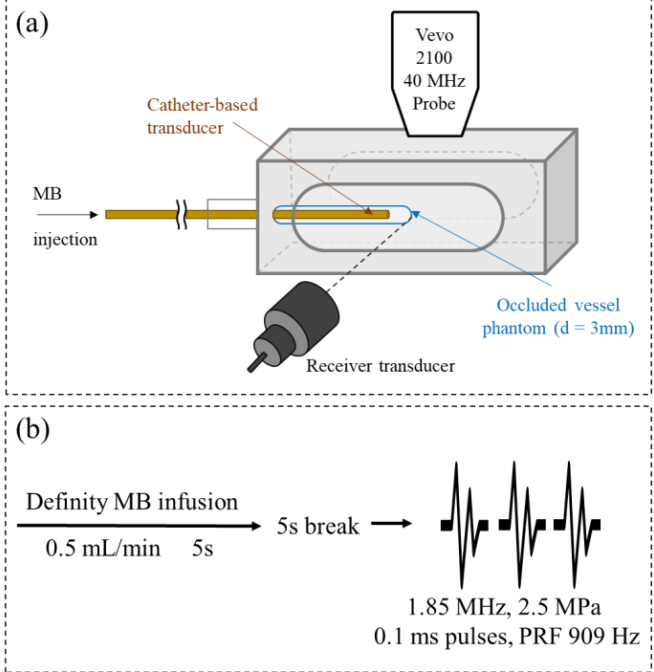

**Fig. 3**. (a) Phantom experiment setup and (b) the pulsing parameters used.

### III. RESULTS

#### A. Cylindrical PZT transducer elements
*1) Simulations*

Fig. 4 shows the simulated frequency domain complex impedance as well as the x- (radial direction) and y- (axial direction) displacements from the broadband excitation of a 2.5 mm cylinder. The graphs show that impedance minima closely align with either x- or y-displacement maxima, corresponding to resonant frequencies of the cylinder. Large y-displacements are consistent with a 'piston' action of the cylinder which is apparent for excitation at 550 kHz and 1850 kHz. In contrast, the resonance frequency implied by the impedance minima at 1095 kHz shows relatively little y-displacement but large x-displacement at that frequency, indicating motion of the cylinder's outer diameter surface.

Simple analytical approximations for resonant modes can be calculated based on the geometry of the cylinder and the speed of sound of the ceramic [58]. For example, the length mode can be calculated as $f_l = c_{PZT}/(2l)$ where $c_{PZT}$ is the speed of sound in the piezoelectric material and $l$ is the cylinder length. For a 2.5 mm length, the calculated value of 578 kHz matches closely with the simulated resonant frequency of 550 kHz. Similarly, the radial mode can be calculated as $f_r = c_{PZT}/(2\pi r)$ where $r$ is the average radius. For an average radius of 0.52 mm, the calculated value is 884 kHz. This does not match the simulated value of 1095 kHz, likely due to the non-negligible wall thickness of the cylinder whereas the analytical approximation assumes a thin ring.

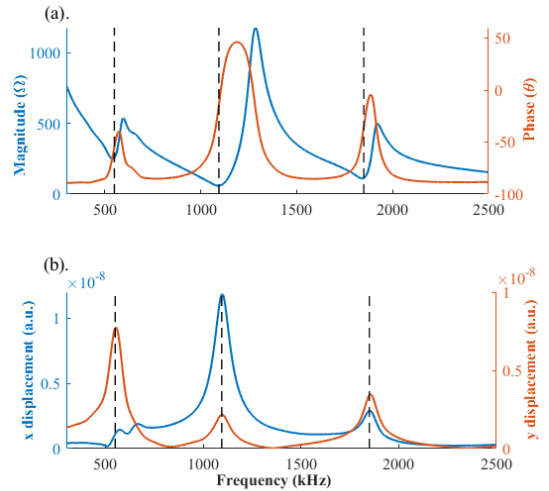

**Fig. 4**. (a) Simulated impedance magnitude and phase for a 2.5 mm transducer. (b) Simulated y-displacement of the front face of the transducer and x-displacement of the cylinder's side surface. Three resonance modes are highlighted at 550 kHz, 1095 kHz, and 1850 kHz with the vertical dashed lines, as characterized by minima in impedance and maxima in displacement.

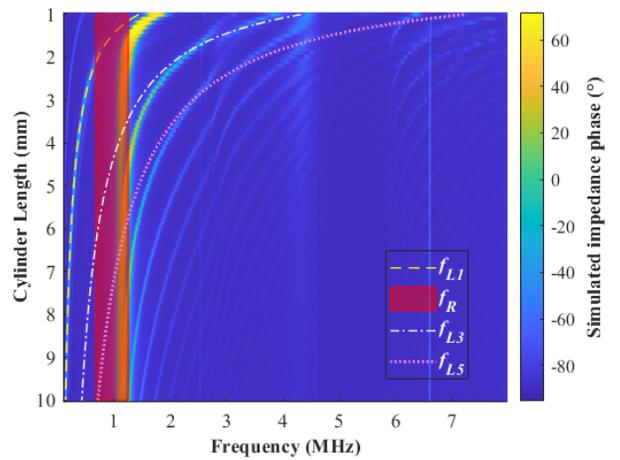

**Fig. 5**. Comparison of simulated impedance phases over a range of frequencies and cylinder lengths. Dashed curves denote analytical approximations for fundamental, 3rd harmonic, and 5th harmonic length mode frequencies while the translucent red region denotes the analytical approximation for the radial mode frequency over the range from the inner to outer radii.

Fig. 5 shows the simulated impedance phases for a range of cylinder lengths. The light blue to yellow regions show maxima in the impedance phase which correspond to different resonant frequencies. This figure highlights how the simple frequency relationships between dimensions and resonant frequencies are not always consistent for complex geometries. For example, the highlighted region at a frequency of around 1100 kHz corresponding to the radial mode extends from 10 mm down to approximately 2 mm, however below this length the frequency of this mode appears to increase despite an unchanging radius, contrary to the analytical approximation. Similarly, the length mode and its harmonics appear to diverge from their analytical approximations as they approach the radial mode frequency. We also note that as the cylinder length is reduced below 1 mm, the impedance rises sharply due to the decreased electrode area.

Excitation at a fixed frequency results in pressure distributions both within the lumen and in outward directions, as shown in Fig. 6(a). Fig. 6(b) shows the pressure along the central axis of the cylinder, which allows for comparison of the three identified frequencies. This figure also highlights how the spatial distribution of pressure varies for different frequencies, and in particular, how the pressure generated internal to the cylinder relative to the pressure directed outwards from the cylinder face can vary significantly depending on resonant mode. In this example, the 3$^{rd}$ harmonic of length mode ($f_{L3}$) at 1.85 MHz shows a similar magnitude of projected pressure relative to within the lumen, whereas the radial mode ($f_R$) at 1.095 kHz shows much higher pressures produced only within the lumen. Meanwhile, the higher order harmonic of length mode is capable of projecting higher forward-looking pressure compared to the fundamental length mode ($f_{L1}$) at 550 kHz.

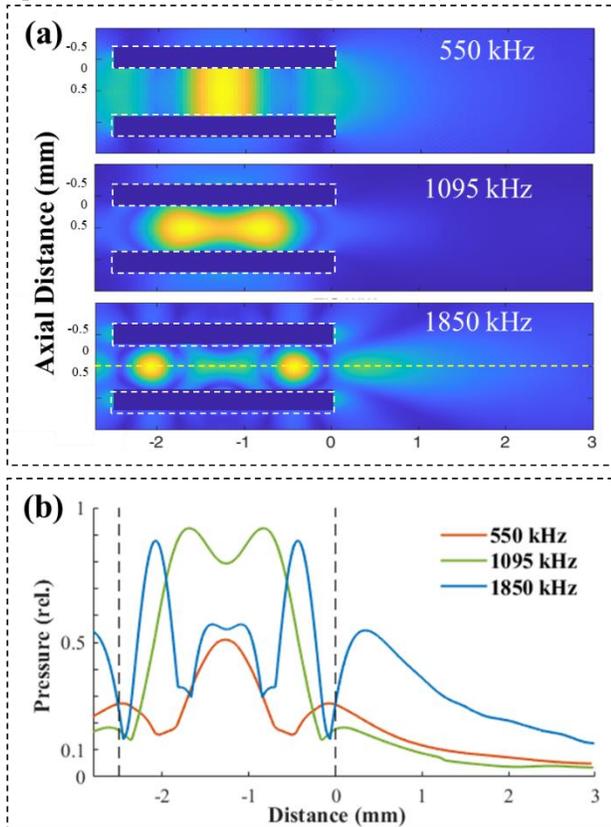

**Fig. 6.** (a) Simulated emitted ultrasound pressure fields for a 2.5 mm transducer excited at 550 kHz, 1095 kHz, and 1850 kHz. The transducer outline is highlighted with white dashed lines. (b) Corresponding comparison of pressure along the central axis (yellow dashed line shown in (a)), with the vertical black dashed lines denoting the front and back faces of the 2.5 mm PZT cylinder.

*2) Experimental*

In Fig. 7, the maximum internal and external pressure at the length mode and length mode harmonic ($f_{L1}$ and $f_{L3}$), as well as the radial mode ($f_R$) of the transducers for 2 to 4 mm lengths are compared. For the application of catheter-based microbubble mediated therapies, the primary considerations for the selected length are (1) maximizing forward looking pressure, (2) minimizing internal pressure, (3) favoring lower frequencies for microbubble excitation, and (4) minimizing length to allow for mechanical flexibility of the distal tip. Together with experimental on-axis pressure profiles (Supplementary Fig. 4), these results show that a length of 2.5 mm operated at the 3$^{rd}$ harmonic produced a high projected external pressure over a significant distance from the face while also having a relatively low pressure internal to the lumen. This configuration was used for the subsequent catheter construction.

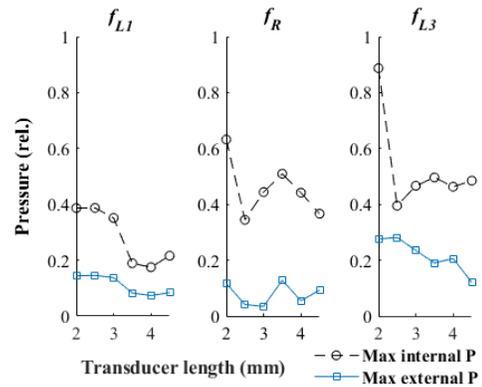

**Fig. 7.** Experimentally measured maximum internal and external pressure along the central axis of the transducers with lengths from 2 to 4.5 mm at selected modes with an applied voltage of 20 V$_{pp}$.

### B. Constructed catheter-based transducers

*1) Forward-looking pressure projection*

The effects of air backing and epoxy matching layers was assessed by comparing the pressure output measured at $f_{L3}$ of a bare 2.5 mm PZT element, a catheter-based transducer without a matching layer, and a prototype with a λ/4 epoxy matching layer. The measured electrical impedance and corresponding $f_{L3}$ are shown in Supplementary Fig. 5. The addition of backing and matching layers indicates a slight damping in impedance as well as a shift in the frequency of resonant modes compared to the bare element.

The external pressure profiles for the aforementioned three transducers are shown in Fig. 8(a) and they display substantial similarities in spatial pattern. Under the same driving voltage, air backing increases the external pressure projection. The addition of the matching layer shows further improved increased pressures (Fig. 8(b)). Four constructed catheters, one with only a backing layer and three with both a backing and a matching layer were tested. The on-axis pressure at an offset of 0.5 mm from the front face, as a function of driving voltage, for the four constructed catheters and a bare PZT element are compared in Fig. 9(a). The bare element shows an earlier plateau of the driving voltage-dependent pressure output. Catheters with both backing and matching layers were able to be driven with a voltage up to ~300 $V_{pp}$, where nearly 3 MPa was achieved at a 0.5 mm offset. At 280 $V_{pp}$, the axial pressure profile delivered in the forward direction by all three catheters having both a backing and a matching layer showed minimal difference (Fig. 9(b)).

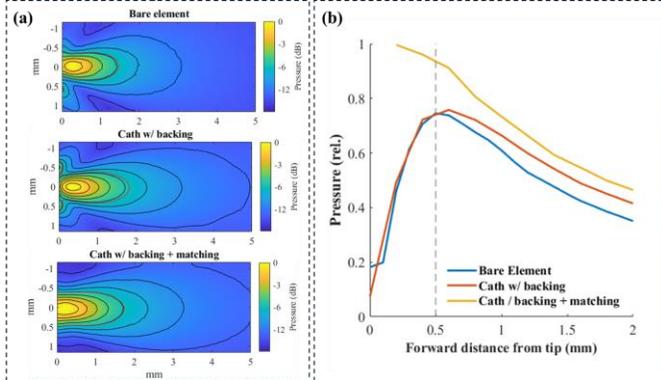

**Fig. 8**. (a) Experimentally measured peak negative pressure profiles and (b) pressures along the central axis for a bare PZT element, a catheter with a backing layer, and a catheter with both a backing and a matching layer, under the same driving conditions and with electrical impedance matching.

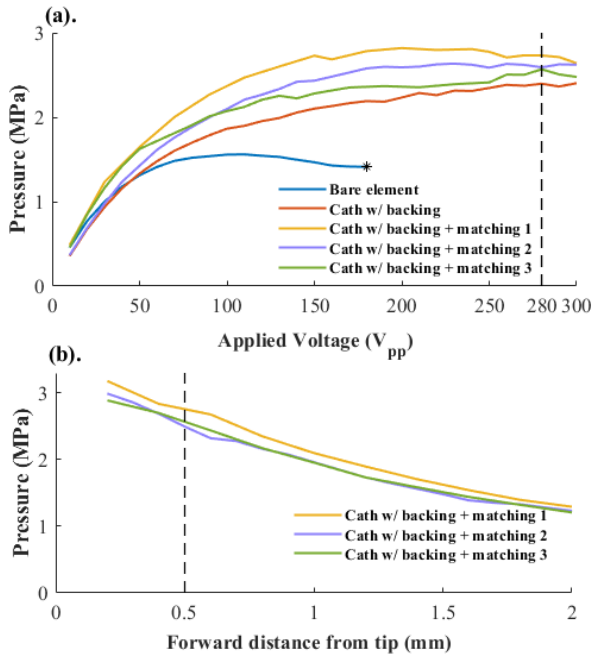

**Fig. 9**. (a) External peak negative pressure at a distance of 0.5 mm away from the front face as a function of applied voltage. The '*' indicates the point that the bare element was observed to crack. (b) The pressure delivered along the central axis at 280 $V_{pp}$ as a function of forward distance from the front face for three catheters with backing and matching layers.

*2) Lateral pressure projection*

Fig. 10 shows the measured lateral pressures for both a bare element and a catheter. The assembled catheter has a reduced lateral pressure profile, which is attributed to the presence of the jacket.

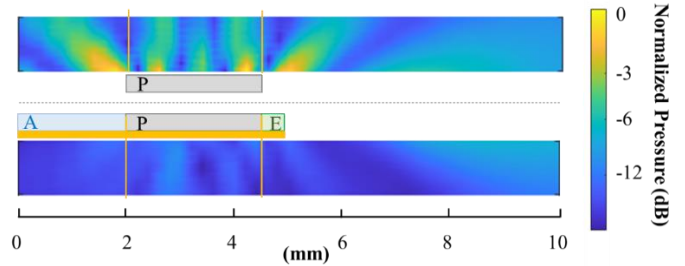

**Fig. 10**. Experimentally measured lateral pressure of a bare PZT element (top) and the constructed catheter transducer with a backing and a matching layer (bottom) at the same driving voltage (both with impedance matched to 50Ω (± 3 Ω) and 0° (± 5°)). A: Air backing layer. P: PZT cylinder. E: Epoxy matching layer.

### C. Vessel phantom demonstration

In the vessel phantom experiments seen in Fig. 11, the catheter-based transducer was placed 1.5 mm away from the occlusion surface. 20 pulses (0.1 ms) were sent with a PRF of 909 Hz and a frequency of 1.85 MHz. The pressures produced ranged from 2.5 MPa at the surface of the transducer to 1.4 MPa at the occlusion surface (1.5 mm offset; peak negative pressures). The frames show the ejection of microbubbles from the catheter tip, followed by sonication and destruction of the microbubbles. Fig 12 shows spectra captured from the receiver transducer during sonication, which exhibit high levels of broadband noise that decrease in amplitude with subsequent pulses, coinciding with the observed destruction of microbubbles.

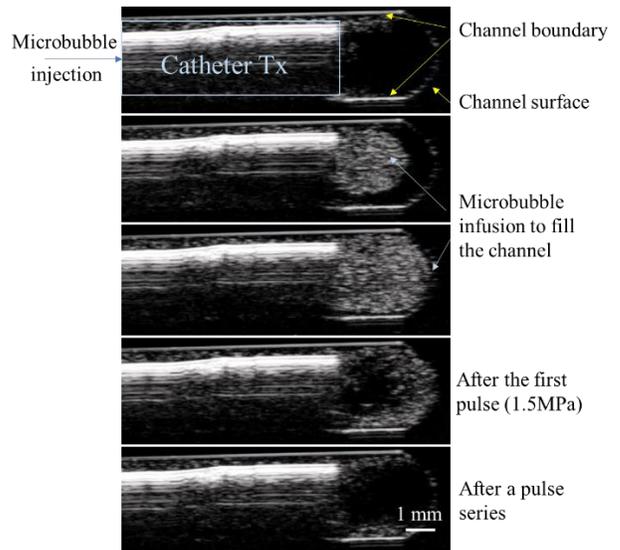

**Fig. 11**. Ultrasound imaging of the infused microbubble cloud translating forward under low-amplitude pulses and accumulating at the occlusion surface. The following high-amplitude disruption pulses induced localized cavitation of the microbubbles.

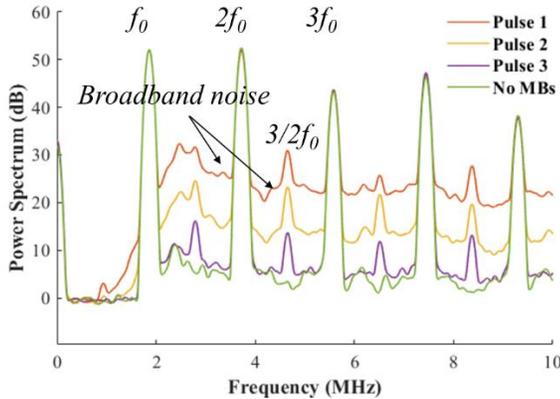

**Fig 12**. Cavitation spectra captured in the first 3 pulses at the channel boundary shown in **Fig. 11**.

## IV. Discussion

Collectively, these results demonstrate the ability of mm-scale hollow cylindrical transducers to produce high pressures directed in a forward-facing direction as well as their ability to be incorporated into catheter assemblies suitable for treating large vessel occlusions. Initial vessel phantom experiments demonstrated the ability of the catheter assembly to allow for injection of microbubbles followed by sonication and inertial cavitation of the microbubbles.

The catheter design and construction were informed by simulations and experiments related to the behavior of the hollow cylindrical transducers. We see from the simulated impedance curves in Fig. 5 that certain modes remain static with varying length while others shift. The modes which are related to the diameter or wall thickness are less affected by changes in cylinder length until the aspect ratio becomes sufficiently small. The electro-mechanical behavior becomes increasingly complex when the dimensions for cylinder length, diameter, and thickness create overlapping vibrational modes.

Simulated impedances and pressure profiles are in broad agreement with experimentally measured impedances and pressure maps, with divergences likely due to differences in material properties and the influence of the physical connection method. The effect of the physical connection method on the impedance was seen with the lightly tensioned spring wire connection showing two impedance minima around the radial mode, while simulations and the assembled catheter construction method using conductive foil and silver epoxy showed only a single minimum (Supplementary Fig. 5).

The selected transducer element length of 2.5 mm and driving mode of $f_{L3}$ was based on considerations of maximizing the projected external pressure field, minimizing the pressures produced within the lumen, ensuring an excitation frequency within a range suitable for microbubble mediated therapies, and being of a size that is compatible with navigating potentially tortuous vessels in a catheter application. Lengths above 2.5 mm were not favored primarily due to lower forward projected pressures and the inherent reduced flexibility of a catheter assembly. The overlap in resonant modes likely contributes to the reduced pressures seen for $f_{L3}$ at lengths above 3.5 mm, where the resonant mode frequencies approach that of the radial mode. Lengths of 2 mm and below were considered, however these had similar or degraded forward projected pressures as well as higher pressures produced internally within the lumen, which may increase the risks of damage to the internal structures of the catheter at higher driving voltages. In addition, the higher frequencies associated with shorter lengths were less desirable from a bubble stimulation perspective.

In addition to validating the ability of cylindrical transducers to form the active element for catheter-based ultrasound therapies, the assembled structures also conferred advantages related to the generation of ultrasound in the proposed therapeutic context. For instance, the outer jacket and inner liner enabled air backing behind the proximal face of the transducer which likely contributes to the increased pressures seen for the assembled catheter relative to the cylindrical element on its own. The addition of an epoxy matching layer further increased the forward projected pressure levels. Furthermore, the lateral pressures emitted radially outwards from the cylindrical surface were reduced, which is preferable from a safety perspective due to reducing the possibilities of unintended damage to surrounding vessel walls. Finally, the assembled catheters were able to be driven at higher voltages and thus produce higher pressures before failing, compared to the bare elements.

Four assembled catheters were constructed, three of which included a matching layer. These three catheters showed broadly similar results with respect to the magnitudes and spatial distribution of the pressure fields produced. The differences seen may be attributable to variability in the transducer elements [59] and the manual assembly process.

Preliminary vessel phantom experiments illustrated the ability to infuse microbubbles through the catheter lumen and induce their destruction forward of the transducer face. Visual ultrasound imaging and spectral recordings confirm the inertial cavitation of the microbubbles. These experiments highlight the potential for treating vascular occlusions such as thrombus or CTOs.

The capability of delivering peak negative pressures of >2.5 MPa at a distance of 0.5 mm from the transducer face are within the range of what has shown effectiveness in previous work on microbubble mediated sonothrombolysis and preclinical CTO therapy. For the presented pilot work, a single pump control and pulsing sequence was chosen to demonstrate viability, however optimization of these parameters could enhance the treatment of vascular occlusions.

## V. Conclusions

We have developed a novel forward looking therapeutic ultrasound catheter based on the use of a hollow cylindrical transducer. The compact design approach permits the introduction of a guidewire, microbubbles, and lytic agents in close proximity to a vascular occlusion. This approach has relevance to both thrombotic and chronic total occlusions.

## References

[1] J. Lin, Y. Chen, N. Jiang, Z. Li, and S. Xu, "Burden of Peripheral Artery Disease and Its Attributable Risk

Factors in 204 Countries and Territories From 1990 to 2019," *Front Cardiovasc Med*, vol. 9, Apr. 2022, doi: 10.3389/fcvm.2022.868370.

[2] V. F. Tapson, "Acute Pulmonary Embolism," *The New England Journal of Medicine*, vol. 358, pp. 1037–1052, 2008, Accessed: Aug. 30, 2023. [Online]. Available: www.nejm.org

[3] C. W. Tsao et al., "Heart Disease and Stroke Statistics - 2023 Update: A Report from the American Heart Association," *Circulation*, vol. 147, no. 8, pp. E93–E621, Feb. 2023, doi: 10.1161/CIR.0000000000001123.

[4] M. Goyal et al., "Endovascular thrombectomy after large-vessel ischaemic stroke: A meta-analysis of individual patient data from five randomised trials," *The Lancet*, vol. 387, no. 10029, pp. 1723–1731, Apr. 2016, doi: 10.1016/S0140-6736(16)00163-X.

[5] K. J. Shah and T. L. Roy, "Catheter-Directed Interventions for the Treatment of Lower Extremity Deep Vein Thrombosis," *Life*, vol. 12, p. 1984, 2022, doi: 10.3390/life12121984.

[6] A. K. Sista et al., "Indigo Aspiration System for Treatment of Pulmonary Embolism Results of the EXTRACT-PE Trial," *JACC Cardiovasc Interv*, vol. 14, pp. 319–329, 2021, doi: 10.1016/j.jcin.2020.09.053.

[7] W. J. Powers et al., "Guidelines for the early management of patients with acute ischemic stroke: 2019 update to the 2018 guidelines for the early management of acute ischemic stroke a guideline for healthcare professionals from the American Heart Association/American Stroke Association," *Stroke*, vol. 50, no. 12. Lippincott Williams and Wilkins, pp. E344–E418, Dec. 01, 2019. doi: 10.1161/STR.0000000000000211.

[8] G. W. Stone et al., "Percutaneous recanalization of chronically occluded coronary arteries: A consensus document - Part II," *Circulation*, vol. 112, no. 16, pp. 2530–2537, 2005, doi: 10.1161/CIRCULATIONAHA.105.583716.

[9] R. Jaffe et al., "Natural History of Experimental Arterial Chronic Total Occlusions," *J Am Coll Cardiol*, vol. 53, no. 13, pp. 1148–1158, 2009, doi: 10.1016/j.jacc.2008.09.064.

[10] M. Katsuragawa, H. Fujiwara, M. Miyamae, and S. Sasayama, "Histologic studies in percutaneous transluminal coronary angioplasty for chronic total occlusion: Comparison of tapering and abrupt types of occlusion and short and long occluded segments," *J Am Coll Cardiol*, vol. 21, no. 3, pp. 604–611, 1993, doi: 10.1016/0735-1097(93)90091-E.

[11] G. W. Stone et al., "Percutaneous recanalization of chronically occluded coronary arteries: A consensus document - Part I," *Circulation*, vol. 112, no. 15. pp. 2364–2372, Oct. 11, 2005. doi: 10.1161/CIRCULATIONAHA.104.481283.

[12] D. Dash, "Guidewire crossing techniques in coronary chronic total occlusion intervention: A to Z," *Indian Heart Journal*, vol. 68, no. 3. Elsevier B.V., pp. 410–420, May 01, 2016. doi: 10.1016/j.ihj.2016.02.019.

[13] S. S. Srivatsa et al., "Histologic Correlates of Angiographic Chronic Total Coronary Artery Occlusions," *JACC*, vol. 29, no. 5, pp. 955–63, 1997, doi: 10.1016/S0735-1097(97)00035-1.

[14] P. Fefer et al., "Current perspectives on coronary chronic total occlusions: The Canadian multicenter chronic total occlusions registry," *J Am Coll Cardiol*, vol. 59, no. 11, pp. 991–997, 2012, doi: 10.1016/j.jacc.2011.12.007.

[15] H. J. Colmenarez et al., "Efficacy and Safety of Drug-Eluting Stents in Chronic Total Coronary Occlusion Recanalization. A Systematic Review and Meta-Analysis," *J Am Coll Cardiol*, vol. 55, no. 17, pp. 1854–1866, 2010, doi: 10.1016/j.jacc.2009.12.038.

[16] E. C. Keeley, J. A. Boura, and C. L. Grines, "Primary angioplasty versus intravenous thrombolytic therapy for acute myocardial infarction: A quantitative review of 23 randomised trials," *Lancet*, vol. 361, no. 9351, pp. 13–20, 2003, doi: 10.1016/S0140-6736(03)12113-7.

[17] A. Huqi, D. Morrone, G. Guarini, and M. Marzilli, "Long-term follow-up of elective chronic total coronary occlusion angioplasty: Analysis from the U.K. central cardiac audit database," *J Am Coll Cardiol*, vol. 64, no. 24, pp. 2707–2708, 2014, doi: 10.1016/j.jacc.2014.09.058.

[18] A. R. Galassi, S. D. Tomasello, L. Costanzo, M. B. Campisano, G. Barrano, and C. Tamburino, "Long-term clinical and angiographic results of sirolimus-eluting stent in complex coronary chronic total occlusion revascularization: The SECTOR registry," *J Interv Cardiol*, vol. 24, no. 5, pp. 426–436, 2011, doi: 10.1111/j.1540-8183.2011.00648.x.

[19] D. M. Safley, J. A. Grantham, J. Hatch, P. G. Jones, and J. A. Spertus, "Quality of life benefits of percutaneous coronary intervention for chronic occlusions," *Catheterization and Cardiovascular Interventions*, vol. 84, no. 4, pp. 629–634, 2014, doi: 10.1002/ccd.25303.

[20] H. J. Aparicio et al., *Heart Disease and Stroke Statistics-2021 Update*. 2021. doi: 10.1161/CIR.0000000000000950.

[21] H. C. Wijeysundera et al., "Relationship between initial treatment strategy and quality of life in patients with coronary chronic total occlusions," *EuroIntervention*, vol. 9, no. 10, pp. 1165–1172, Feb. 2014, doi: 10.4244/EIJV9I10A197.

[22] L. P. Hoebers, B. E. Claessen, G. D. Dangas, T. Ramunddal, R. Mehran, and J. P. S. Henriques, "Contemporary overview and clinical perspectives of chronic total occlusions," *Nat. Rev. Cardiol*, vol. 11, pp. 458–469, 2014, doi: 10.1038/nrcardio.2014.74.

[23] M. L. Narducci et al., "Mid-Term Outcome of Ventricular Arrhythmias Catheter Ablation in Patients with Chronic Coronary Total Occlusion Compared to Ischemic and Non-Ischemic Patients," *J Clin Med*,

vol. 11, no. 23, Dec. 2022, doi: 10.3390/jcm11237181.
[24] K. Denby, L. Young, S. Ellis, and J. Khatri, "Antegrade wire escalation in chronic total occlusions: State of the art review," *Cardiovascular Revascularization Medicine*. Elsevier Inc., 2023. doi: 10.1016/j.carrev.2023.06.011.
[25] S. Tummala and A. J. Richardson, "Infrapopliteal Artery Chronic Total Occlusion Crossing Techniques: An Overview for Endovascular Specialists," *Semin Intervent Radiol*, vol. 38, no. 4, pp. 492–499, Oct. 2021, doi: 10.1055/s-0041-1735607.
[26] E. B. Wu *et al.*, "Global Chronic Total Occlusion Crossing Algorithm: JACC State-of-the-Art Review," *Journal of the American College of Cardiology*, vol. 78, no. 8. Elsevier Inc., pp. 840–853, Aug. 24, 2021. doi: 10.1016/j.jacc.2021.05.055.
[27] J. J. Graham *et al.*, "Collagenase to facilitate guidewire crossing in chronic total occlusion PCI—The Total Occlusion Study in Coronary Arteries-5 (TOSCA-5) trial," *Catheterization and Cardiovascular Interventions*, vol. 99, no. 4, pp. 1065–1073, Mar. 2022, doi: 10.1002/ccd.30101.
[28] B. H. Strauss *et al.*, "Collagenase total occlusion-1 (CTO-1) trial: A phase I, dose-escalation, safety study," *Circulation*, vol. 125, no. 3, pp. 522–528, Jan. 2012, doi: 10.1161/CIRCULATIONAHA.111.063198.
[29] A. B. Osherov *et al.*, "Effects of intracoronary collagenase injection in a porcine model: A safety dose-finding study," *J Clin Exp Cardiolog*, vol. 5, no. 7, 2014, doi: 10.4172/2155-9880.1000323.
[30] N. Kucher *et al.*, "Randomized, controlled trial of ultrasound-assisted catheter-directed thrombolysis for acute intermediate-risk pulmonary embolism," *Circulation*, vol. 129, no. 4, pp. 479–486, Jan. 2014, doi: 10.1161/CIRCULATIONAHA.113.005544.
[31] C. K. Holland, S. S. Vaidya, S. Datta, C.-C. Coussios, and G. J. Shaw, "Ultrasound-enhanced tissue plasminogen activator thrombolysis in an in vitro porcine clot model," 2007, doi: 10.1016/j.thromres.2007.07.006.
[32] M. Daffertshofer *et al.*, "Transcranial Low-Frequency Ultrasound-Mediated Thrombolysis in Brain Ischemia Increased Risk of Hemorrhage With Combined Ultrasound and Tissue Plasminogen Activator Results of a Phase II Clinical Trial," 2005, doi: 10.1161/01.STR.0000170707.86793.1a.
[33] A. V Alexandrov *et al.*, "Ultrasound Enhanced Systemic Thrombolysis for Acute Ischemic Stroke," *N Engl J Med*, vol. 21, pp. 2170–2178, 2004, Accessed: Aug. 29, 2023. [Online]. Available: www.nejm.org
[34] S.-L. Yang *et al.*, "Delivery of CD151 by Ultrasound Microbubbles in Rabbit Myocardial Infarction," *Cardiology*, vol. 135, pp. 221–227, 2016, doi: 10.1159/000446639.
[35] B. Petit *et al.*, "In Vitro Sonothrombolysis of Human Blood Clots with BR38 Microbubbles," *Ultrasound Med Biol*, vol. 38, no. 7, pp. 1222–1233, Jul. 2012, doi: 10.1016/j.ultrasmedbio.2012.02.023.
[36] S. Datta, C. C. Coussios, A. Y. Ammi, T. D. Mast, G. M. de Courten-Myers, and C. K. Holland, "Ultrasound-Enhanced Thrombolysis Using Definity® as a Cavitation Nucleation Agent," *Ultrasound Med Biol*, vol. 34, no. 9, pp. 1421–1433, Sep. 2008, doi: 10.1016/J.ULTRASMEDBIO.2008.01.016.
[37] J. Kim, R. M. DeRuiter, L. Goel, Z. Xu, X. Jiang, and P. A. Dayton, "A Comparison of Sonothrombolysis in Aged Clots between Low-Boiling-Point Phase-Change Nanodroplets and Microbubbles of the Same Composition," *Ultrasound Med Biol*, vol. 46, no. 11, pp. 3059–3068, Nov. 2020, doi: 10.1016/J.ULTRASMEDBIO.2020.07.008.
[38] S. Guo *et al.*, "Reduced clot debris size in sonothrombolysis assisted with phase-change nanodroplets," *Ultrason Sonochem*, vol. 54, pp. 183–191, Jun. 2019, doi: 10.1016/J.ULTSONCH.2019.02.001.
[39] D. Pajek, A. Burgess, Y. Huang, and K. Hynynen, "High-Intensity Focused Ultrasound Sonothrombolysis: The Use of Perfluorocarbon Droplets to Achieve Clot Lysis at Reduced Acoustic Power," *Ultrasound Med Biol*, vol. 40, no. 9, pp. 2151–2161, Sep. 2014, doi: 10.1016/J.ULTRASMEDBIO.2014.03.026.
[40] K. B. Bader, S. A. Hendley, and V. Bollen, "Assessment of Collaborative Robot (Cobot)-Assisted Histotripsy for Venous Clot Ablation," *IEEE Trans Biomed Eng*, vol. 68, no. 4, 2021, doi: 10.1109/TBME.2020.3023630.
[41] A. D. Maxwell, C. A. Cain, A. P. Duryea, L. Yuan, H. S. Gurm, and Z. Xu, "Noninvasive Thrombolysis Using Pulsed Ultrasound Cavitation Therapy - Histotripsy", doi: 10.1016/j.ultrasmedbio.2009.07.001.
[42] A. S. Thind *et al.*, "The use of ultrasound-stimulated contrast agents as an adjuvant for collagenase therapy in chronic total occlusions," *EuroIntervention*, vol. 10, no. 4, pp. 484–493, 2014, doi: 10.4244/EIJV10I4A82.
[43] D. E. Goertz *et al.*, "In vivo feasibility study of ultrasound potentiated collagenase therapy of chronic total occlusions," *Ultrasonics*, vol. 54, no. 1, pp. 20–24, 2014, doi: 10.1016/j.ultras.2013.07.014.
[44] M. J. Garcia, "Endovascular Management of Acute Pulmonary Embolism Using the Ultrasound-Enhanced EkoSonic System," 2015, doi: 10.1055/s-0035-1564707.
[45] A. Soltani, K. R. Volz, and D. R. Hansmann, "Effect of modulated ultrasound parameters on ultrasound-induced thrombolysis," *Phys. Med. Biol*, vol. 53, p. 6837, 2008, doi: 10.1088/0031-9155/53/23/012.
[46] L. Goel *et al.*, "Examining the influence of low-dose tissue plasminogen activator on microbubble-mediated forward-viewing intravascular sonothrombolysis," *Ultrasound Med Biol*, vol. 46, pp.

[47] B. Zhang, H. Kim, H. Wu, Y. Gao, and X. Jiang, "Sonothrombolysis with magnetic microbubbles under a rotational magnetic field," 2019, doi: 10.1016/j.ultras.2019.06.004.

[48] J. Kim *et al.*, "Intravascular forward-looking ultrasound transducers for microbubble-mediated sonothrombolysis," *Sci Rep*, vol. 7, no. 1, 2017, doi: 10.1038/s41598-017-03492-4.

[49] L. Goel *et al.*, "Nanodroplet-mediated catheter-directed sonothrombolysis of retracted blood clots," *Microsyst Nanoeng*, vol. 7, no. 1, 2021, doi: 10.1038/s41378-020-00228-9.

[50] R. L. Goldberg and S. W. Smith, "Optimization of signal-to-noise ratio for multilayer PZT transducers," *Ultrason Imaging*, vol. 17, no. 2, pp. 95–113, 1995, doi: 10.1177/016173469501700202.

[51] A. R. Galassi, S. D. Tomasello, L. Costanzo, M. B. Campisano, F. MarzÀ, and C. Tamburino, "Recanalization of complex coronary chronic total occlusions using high-frequency vibrational energy CROSSER catheter as first-line therapy: A single center experience," *J Interv Cardiol*, vol. 23, no. 2, pp. 130–138, Apr. 2010, doi: 10.1111/J.1540-8183.2010.00526.X.

[52] N. Dahdah, A. R. Ibrahim, and A. L. Cannon, "First Recanalization of a Coronary Artery Chronic Total Obstruction in an 11-Year-Old Child with Kawasaki Disease Sequelae Using the CROSSER Catheter", doi: 10.1007/s00246-006-0083-3.

[53] M. S. B. M. C. E. B. M. S. D. Ms. L.-P. R. Ms. M. B. P. P. G. M. Andrew Benko, "Novel Crossing System for Chronic Total Occlusion Recanalization: First-in-Man Experience With the SoundBite Crossing System," *Journal of Invasive Cardiology*, vol. 29, no. 2, Jan. 2017.

[54] J. Song and K. Hynynen, "Feasibility of Using Lateral Mode Coupling Method for a Large Scale Ultrasound Phased Array for Noninvasive Transcranial Therapy", doi: 10.1109/TBME.2009.2028739.

[55] K. Hynynen *et al.*, "Cylindrical Ultrasonic Transducers for Cardiac Catheter Ablation," 1997.

[56] K. Hynynen and J. Yin, "Lateral mode coupling to reduce the electrical impedance of small elements required for high power ultrasound therapy phased arrays," *IEEE Trans Ultrason Ferroelectr Freq Control*, vol. 56, no. 3, pp. 557–564, 2009, doi: 10.1109/TUFFC.2009.1072.

[57] J. Lee Rena, M. Buchanan, L. J. Kleine, and K. Hynynen, "Arrays of multielement ultrasound applicators for interstitial hyperthermia," *IEEE Trans Biomed Eng*, vol. 46, no. 7, pp. 880–890, 1999, doi: 10.1109/10.771202.

[58] J. Song and K. Hynynen, "Feasibility of using lateral mode coupling method for a large scale ultrasound phased array for noninvasive transcranial therapy," *IEEE Trans Biomed Eng*, vol. 57, no. 1, pp. 124–133, Jan. 2010, doi: 10.1109/TBME.2009.2028739.

[59] M. Stewart and M. G. Cain, "Direct Piezoelectric Measurement: The Berlincourt Method," in *Characterisation of Ferroelectric Bulk Materials and Thin Films*, Springer Netherlands, 2014, p. 62. doi: 10.1007/978-1-4020-9311-1_3.

**SUPPLEMENTARY FIGURES**

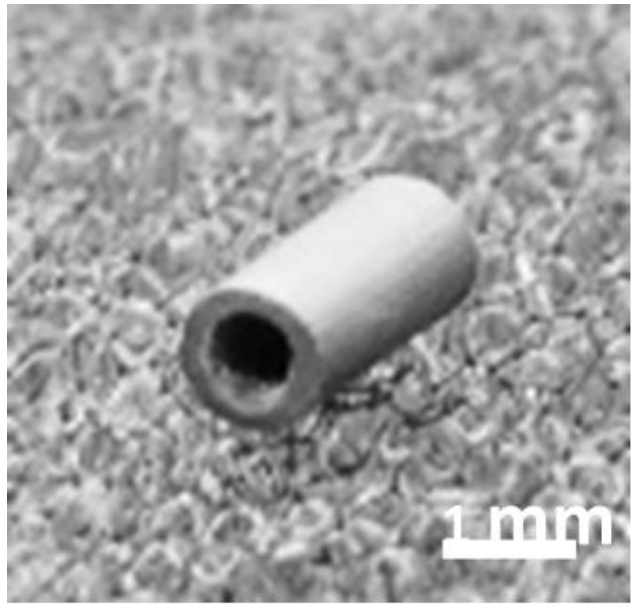

**Supplementary Fig. 1**. An example of the piezoelectric cylinders used as the active element.

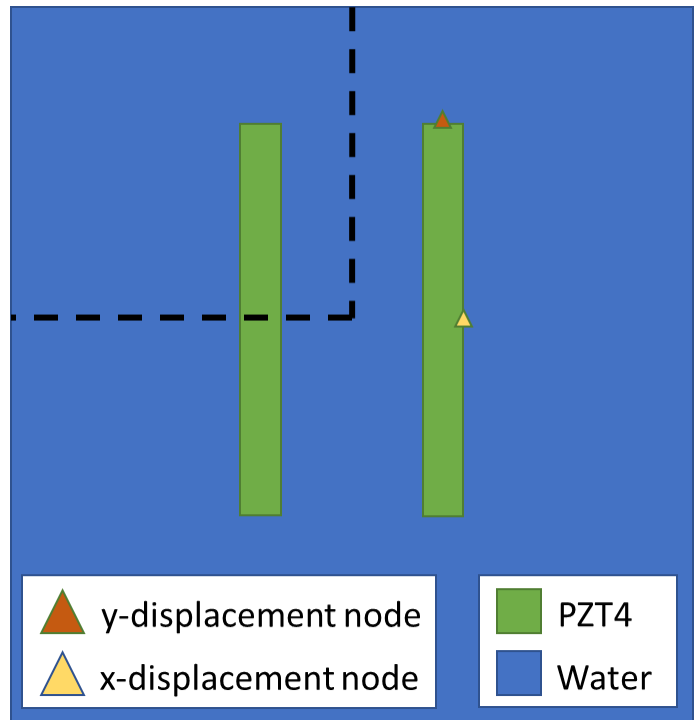

**Supplementary Fig. 2**. Axi-symmetric model used in OnScale simulations. Electrodes are along the inner (signal) and outer (ground) diameters. Dashed black line denotes axis of symmetry, with only ¼ of the transducer modeled as an axisymmetric rotated block.

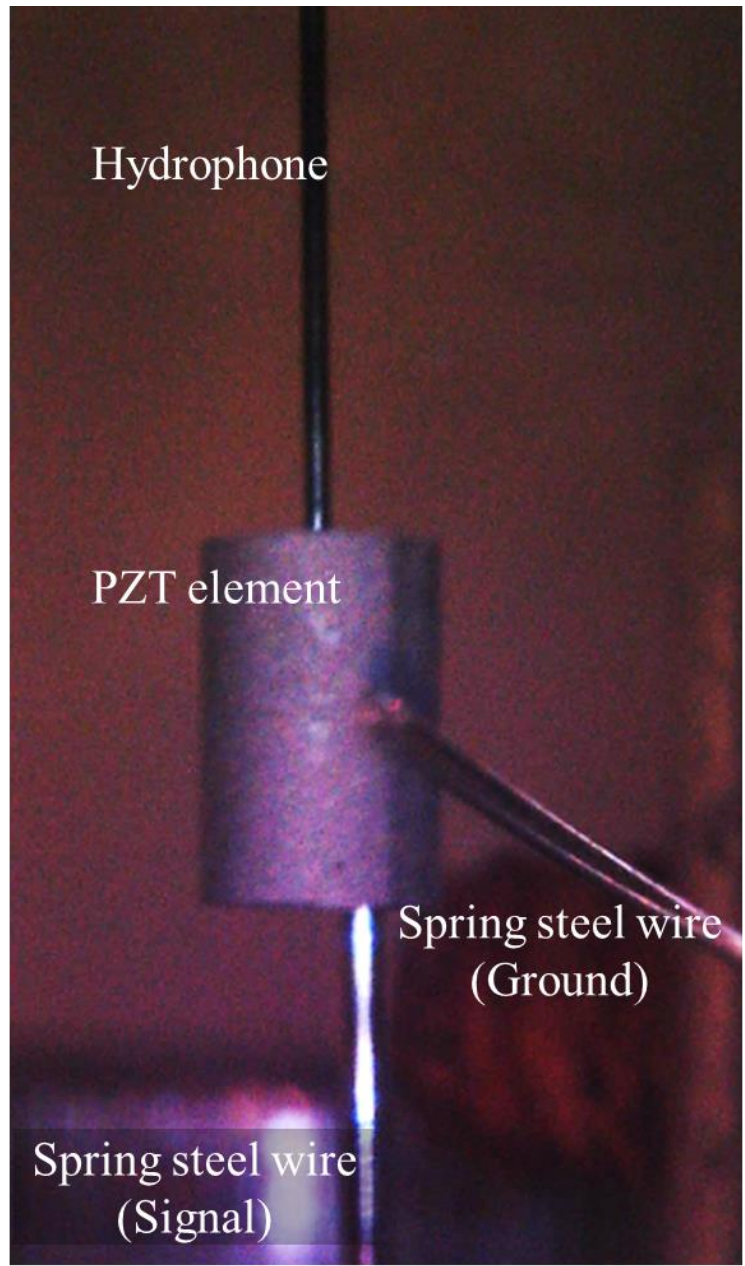

**Supplementary Fig. 3.** Attachment of element on tube (signal connection) with outer ground spring connection. Hydrophone is seen at the top.

(a).

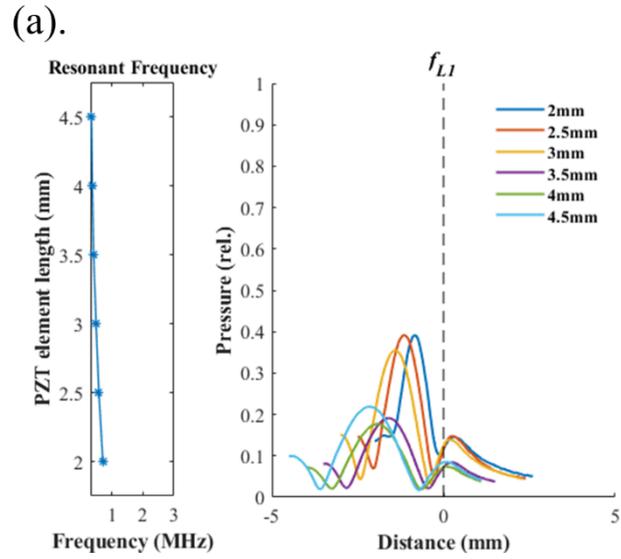

(b).

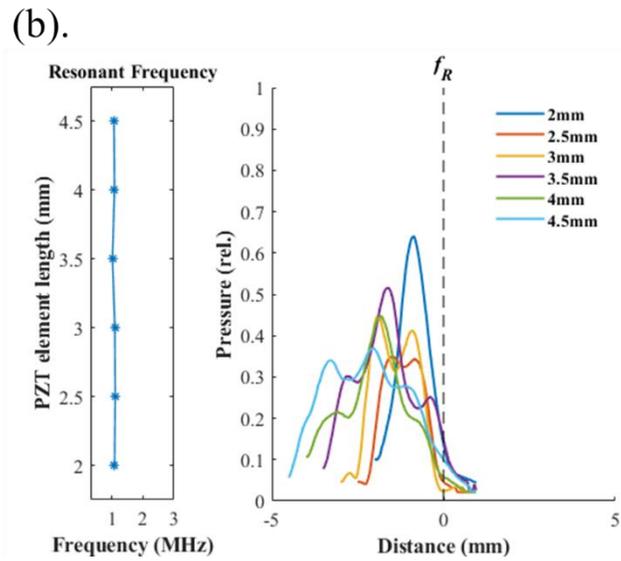

(c).

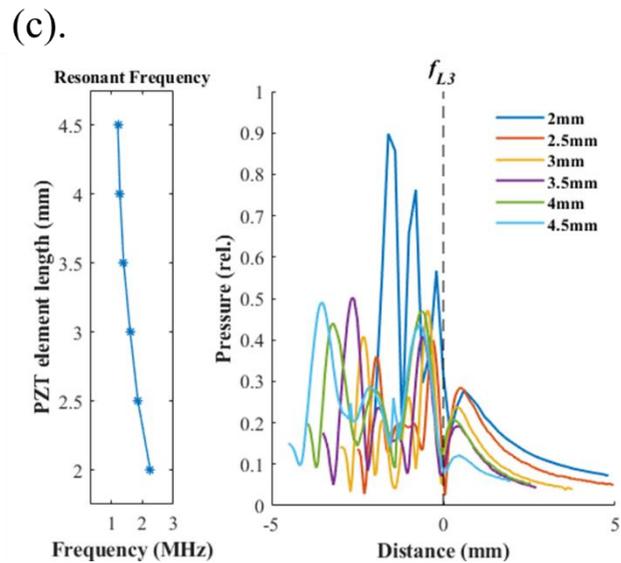

**Supplementary Fig. 4**. Experimental external axial pressure of bare PZT elements at 2 – 4.5 mm at (a). fL1, (b). fR, and (c). fL3 with resonant frequencies plotted on the left panels respectively.

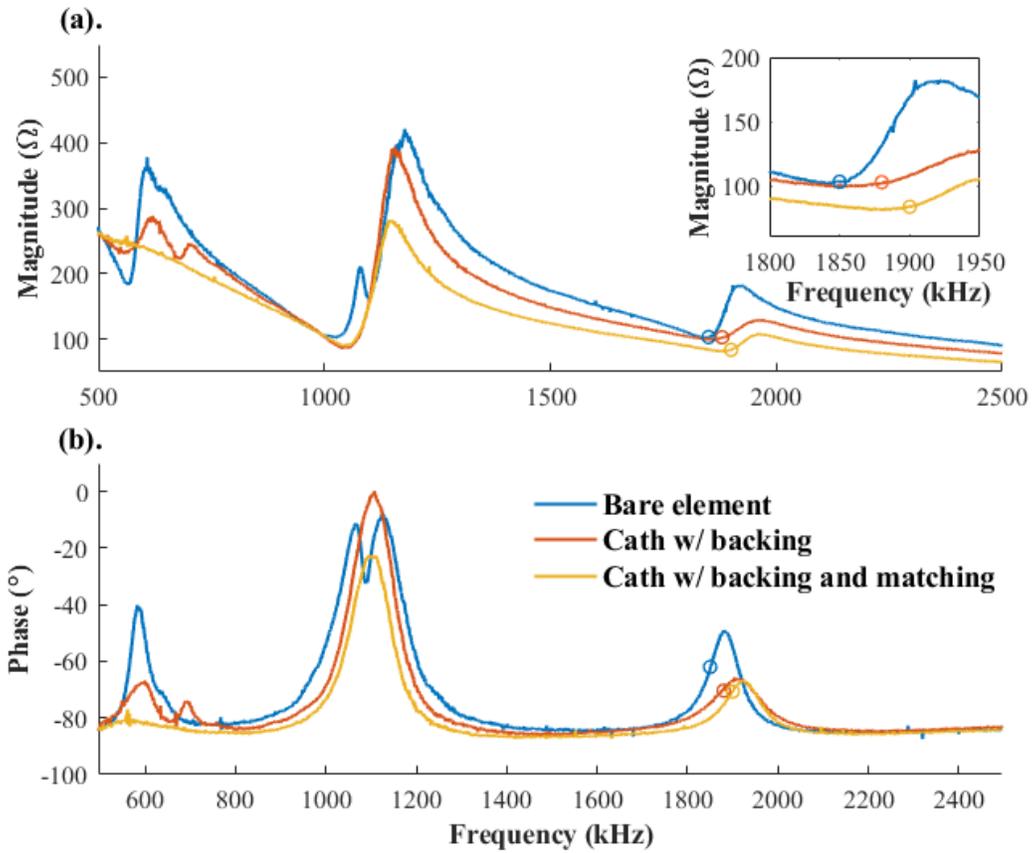

**Supplementary Fig. 5**. Comparison of measured electrical impedances of a bare PZT element, a catheter with a backing layer, and a catheter with a backing and a matching layer. (a). The resonant frequency of fL3 for individual transducer (circles) was determined based on the